\documentclass[aps,10pt,twocolumn,pre,superscriptaddress,notitlepage]{revtex4-2}
\usepackage[utf8]{inputenc}
\usepackage[T1]{fontenc}
\usepackage{graphicx}
\usepackage{booktabs}
\usepackage[most]{tcolorbox}  
\usepackage{enumitem}
\usepackage{booktabs}
\usepackage{ragged2e}
\renewcommand{\arraystretch}{1.08}
\definecolor{natureboxbg}{RGB}{255, 250, 240}

\usepackage{draftwatermark}
\SetWatermarkText{DRAFT}
\SetWatermarkScale{4}
\SetWatermarkLightness{0.93}

\usepackage{microtype}
\usepackage{graphicx}
\usepackage{booktabs}
\usepackage{multirow}
\usepackage{amsmath}
\usepackage{hyperref}
\usepackage{url}
\usepackage{lineno}

\usepackage{booktabs}
\usepackage[most]{tcolorbox}
\usepackage{xurl}

\usepackage[most]{tcolorbox}
\usepackage{xcolor}
\usepackage{enumitem}

\usepackage{xurl}
 
\DeclareRobustCommand{\field}[1]{\nolinkurl{#1}}

\tcbset{
  auditbox/.style={
    enhanced,
    colback=gray!3,
    colframe=black!18,
    boxrule=0.45pt,
    arc=1.2mm,
    left=1.5mm,
    right=1.5mm,
    top=1.2mm,
    bottom=1.2mm,
    before skip=4pt,
    after skip=4pt
  }
}

\definecolor{darkblue}{rgb}{0,0,0.5}
\hypersetup{colorlinks=true,citecolor=darkblue,linkcolor=darkblue,urlcolor=darkblue}

\definecolor{natureboxbg}{RGB}{245,242,226} 

\newcounter{natbox}

\newcommand{\boxheader}{%
  \noindent
  \colorbox{natureboxbg}{%
    \parbox{\dimexpr\columnwidth-2\fboxsep\relax}{%
      \textcolor{black}{\bfseries\normalsize\strut\hspace{3pt} BOX~\thenatbox}%
    }%
  }%
  \par\vspace{-1.0ex}\nointerlineskip
}

\newtcolorbox{natboxinner}[2][]{%
  enhanced, breakable=false, width=\columnwidth,
  colframe=black, boxrule=0.4pt, arc=0mm,
  colback=natureboxbg, left=6pt,right=6pt,top=6pt,bottom=6pt, boxsep=3pt,
  fontupper=\footnotesize, before skip=0pt, after skip=10pt,
  title={#2}, title filled, colbacktitle=white, coltitle=black,
  fonttitle=\bfseries\normalsize,
  boxed title style={colframe=black, boxrule=0.4pt, arc=0mm, left=6pt,right=6pt,top=6pt,bottom=5pt},
  titlerule=0.4pt, titlerule style={black}, bottomtitle=3pt,
  attach boxed title to top left={yshift=-2mm, xshift=0pt},
  #1
}

\setlist[itemize]{leftmargin=*, itemsep=1pt, topsep=2pt, parsep=0pt}

\usepackage{xcolor}
 
\begin{document}

\begin{center}
\emph{Working paper --- subject to revision.  \\}
\end{center}  

\title{How You Ask Shapes What You Get: \\ Auditing Breast-Cancer Misinformation in TikTok Search}

\author{Pooriya Jamie}
\email{pjamie@ucla.edu}
\affiliation{Department of Communication, University of California, Los Angeles, Los Angeles, CA, USA}

\author{Homa Hosseinmardi}
\email{homahm@ucla.edu}
\affiliation{Department of Communication, University of California, Los Angeles, Los Angeles, CA, USA}

\author{Rezvaneh Rezapour}
\affiliation{Department of Information Science, Drexel University, Pennsylvania, USA}

\author{Aria Pessianzadeh}
\affiliation{Department of Information Science, Drexel University, Pennsylvania, USA}

\author{Patricia A. Ganz}
\affiliation{Department of Health Policy and Management, UCLA Fielding School of Public Health, University of California, Los Angeles, Los Angeles, CA, USA}
\affiliation{Division of Hematology/Oncology, Department of Medicine, David Geffen School of Medicine at UCLA, University of California, Los Angeles, Los Angeles, CA, USA}

\author{Amir Ghasemian}
\email{amirgh@ucla.edu}
\affiliation{Department of Communication, University of California, Los Angeles, Los Angeles, CA, USA}


\begin{abstract}
Millions of people use TikTok to seek health information, yet little is known about how users’ search queries shape exposure to health misinformation. Whereas prior algorithm audits have focused primarily on recommendation feeds, we examine TikTok’s search system, where users explicitly express their information needs through query formulation. We conduct a controlled sock-puppet audit of TikTok Search using 30 fresh accounts assigned to six experimental conditions spanning three information-seeking framings (Medical Information, Alternative Medicine, and Peer Narrative) and two breast-cancer contexts (Symptom Noticing and Active Treatment). Across 9,020 retrieved videos, annotated using a validated vision-language model pipeline, we find that query framing is strongly associated with misinformation exposure. Alternative Medicine queries returned misinformation in 54.1\% of cancer-relevant results within the Symptom Noticing context and 53.5\% within the Active Treatment context---8.6× and 7.6× higher, respectively, than clinically framed Medical Information queries. Even Medical Information queries returned measurable levels of possible misinformation (6.3\%–7.1\%), suggesting that explicit medical intent alone does not eliminate exposure. Moreover, for Alternative Medicine queries, possible misinformation appeared throughout the ranked search results rather than only near the top, showing that exposure is not confined to the highest-ranked results. Videos labeled as misinformation were also substantially more likely to contain comments promoting unsupported treatments or anti-standard-care views. These findings demonstrate that search query framing plays a central role in shaping misinformation exposure on TikTok and highlight the importance of auditing query-driven search systems alongside recommendation algorithms.  

\end{abstract}


\flushbottom
\maketitle

\section{Introduction}
\vspace{-2mm}
\label{sec:intro}

People increasingly turn to social media platforms for health information~\citep{Chen2025,suarez2021,chou2018addressing}, including when making sense of symptoms, diagnoses, and treatment decisions. Cancer misinformation on these platforms can attract substantial engagement~\citep{johnson2022}. This is of special concern as research has shown misinformation spreads faster and farther than accurate content~\citep{vosoughi2018}, and documented TikTok cancer content documents
substantial misinformation,  poor quality information, and distinct
alternative-health discourse patterns~\citep{morton2023,muenster2024,xu2021}.

Most existing audits of TikTok focus on algorithmically curated recommendation feeds such as the \emph{For You Page} (FYP), where exposure accumulates from users' behavioral signals over time. In contrast, TikTok Search has received relatively little attention. Search differs fundamentally from feed recommendation because users explicitly communicate their information needs through typed queries. Consequently, exposure depends not only on the ranking algorithm but also on how users formulate those queries.
This distinction is particularly important for health information seeking. A patient searching for \textit{``natural ways to shrink a breast lump''} expresses a fundamentally different information-seeking orientation from one searching for \textit{``breast lump biopsy procedure.''} Whether these different query framings systematically retrieve different levels of health misinformation remains an open empirical question with important implications for platform governance and public health.

We address this question through a controlled sock-puppet audit of TikTok Search. We create 30 fresh TikTok accounts assigned to six experimental conditions spanning three information-seeking framings (Medical Information, Peer Narrative, and Alternative Medicine) and two breast-cancer contexts (Symptom Noticing and Active Treatment). Search queries are generated based on discussions from the \textit{r/breastcancer} community, and the resulting search sessions yield  9,020 videos, of which 7,199 were cancer-relevant. A prompted vision-language model (VLM) then labels the collected videos for cancer relevance and potential misinformation. We organize the study around four research questions:

\begin{itemize}
\item \textbf{RQ1 (Framing).} Within a fixed search context, how do
  misinformation-exposure rates in TikTok search results differ across search query framings?
  \item \textbf{RQ2 (Context).} Does exposure differ between the
  Symptom Noticing and Active Treatment search contexts?
  \item \textbf{RQ3 (Rank).} How is misinformation distributed across the ranked result list?
  \item \textbf{RQ4 (Comments).} How does the content of comments differ between videos labeled as misinformation and those not?
  
\end{itemize}

Our audit reveals that query framing is strongly associated with misinformation exposure. Among cancer-relevant results, Alternative Medicine queries returned possible misinformation at 8.6$\times$ the rate of Medical Information searches during Symptom Noticing (54.1\% versus 6.3\%) and 7.6$\times$ the rate during Active Treatment (53.5\% versus 7.1\%). Moreover, misinformation is distributed throughout the ranked search results rather than appearing only near the top. Even Medical Information queries returned measurable levels of misinformation, suggesting that explicit medical intent alone does not eliminate exposure to harmful content.
Beyond these empirical findings, our work demonstrates how recent LLMs and VLMs can support scalable audits of search systems. By combining LLM-generated queries with VLM-assisted annotation, we present a practical framework for measuring misinformation exposure across thousands of search results.

Specifically, this paper makes four contributions:

\begin{itemize}
\item We present the first large-scale audit of misinformation exposure in TikTok Search for cancer information.

\item We quantify how different information-seeking framings are associated with substantially different misinformation exposure.

\item We characterize how misinformation is distributed across search rankings and compare exposure between Symptom Noticing and Active Treatment contexts.

\item We demonstrate an LLM/VLM-based pipeline for scalable auditing of patient-facing search systems.
\end{itemize}

\vspace{-12pt}
\section{Related work}
\vspace{-10pt}

\textbf{Algorithmic auditing and sock-puppet methods.}
Online platforms have come under increasing scrutiny for a range of societal outcomes, including filter bubbles and radicalization pathways~\citep{haroon2023,ribeiro2020}, partisan bias in Google Search~\citep{robertson2018}, and vaccine misinformation on e-commerce platforms~\citep{juneja2021}. Methods for studying these outcomes range from observational analysis of platform traces~\citep{ribeiro2020} to audits of real-user browsing data~\citep{robertson2018,hosseinmardi2021examining}. Among them, sock-puppet audits create accounts that interact with a platform under predefined conditions while monitoring the outcomes of algorithms~\citep{sandvig2014,ibrahim2026systematic}, and counterfactual-agent designs extend this approach to estimate the causal role of recommender systems in content exposure~\citep{Hosseinmardi2024}. 

Within this tradition, search has received less attention than feeds. Prior work on YouTube search shows that viewing history can shift the share of misinformation among returned results, particularly for vaccine-related queries~\citep{hussein2020}. On TikTok, recent sock-puppet audits focus primarily on the For-You Page, where recommendations adapt to behavioral signals accumulated over prior sessions~\citep{jamie2026}. We instead audit TikTok's query-driven search ranking. Search is also a ranking system, but one in which users state their information need explicitly at each interaction. We operationalize information-seeking differences through LLM-generated query vocabulary grounded in cancer-community Reddit titles, while holding the browser automation and data-collection procedures constant across experimental conditions.

\textbf{Health misinformation on social media.}
Health misinformation is a recurring problem on social media, although its prevalence varies across platforms and health domains: systematic reviews estimate that 20--87\% of studied health-related social media content contains misinformation~\citep{suarez2021}. Cancer is a particularly high-stakes domain, as cancer misinformation attracts substantial engagement and can reach patients at moments of high uncertainty about symptoms, diagnosis, or treatment~\citep{johnson2022}. On TikTok, prior work has documented inaccurate or low-quality cancer content in gynecologic~\citep{morton2023} and prostate~\citep{xu2021} cancer, identified visually distinct alternative-health cancer communities~\citep{muenster2024}, and shown that health misinformation draws user engagement~\citep{baghdadi2023}. These studies characterize the content available on the platform rather than what users are
exposed to upon various search queries, and none addresses breast cancer.

\textbf{LLMs and VLMs for misinformation auditing.}
Automated fact-checking for public-health claims commonly pairs a claim with retrieved external evidence, sometimes with an emphasis on explainability~\citep{kotonya2020}. For video content, multimodal approaches combine textual, visual, and social signals~\citep{akhtar2023}. Closest to our setting, \citet{qi2023} introduce a short-video fake-news benchmark that integrates frame-level visual features, on-screen text, and comment context, making it the nearest structural analog to our screenshot-based annotation pipeline. They train a supervised model on a labeled corpus, whereas we prompt a general-purpose VLM zero-shot. This follows a growing line of work using LLMs as scalable annotators for social-science measurement~\citep{liyanage2024}. We use GPT-5.4-nano, a prompted vision-language model, to classify video screenshots as an annotation step in our audit pipeline, and we validate its outputs against human labels.

\vspace{-12pt}
\section{Methodology}
\vspace{-2mm}
\label{sec:methods}

\subsection{Study Design}
\vspace{-10pt}

We conduct a controlled sock-puppet audit of TikTok Search using a $3 \times 2$
factorial design that crosses three \emph{query framings} with two breast-cancer
\emph{search contexts}, yielding six experimental conditions.

\begin{figure}
    \centering
    \includegraphics[width=\linewidth]{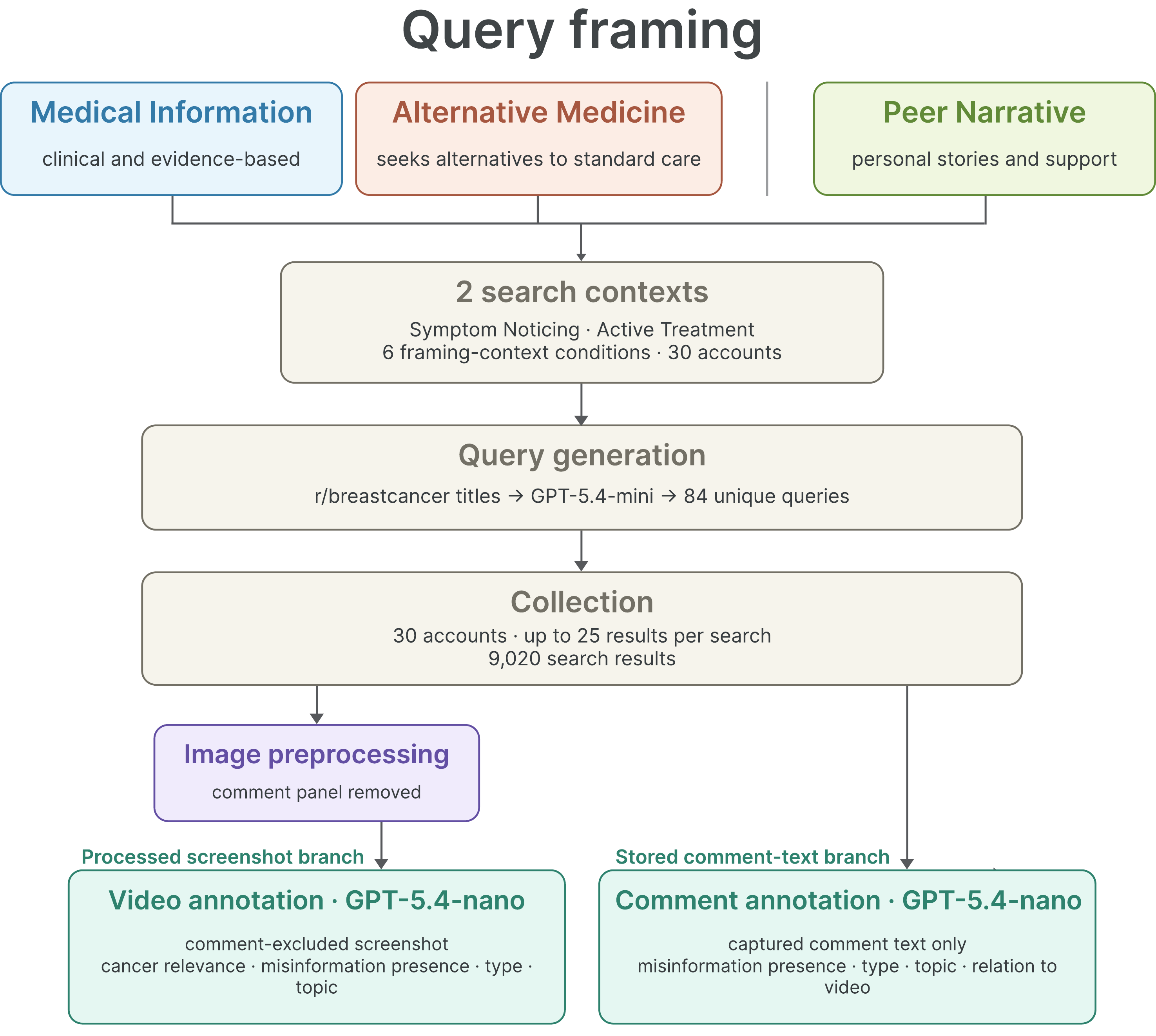}
    \caption{Overview of the methodology. A $3\times2$ design crosses three query
framings (Medical Information, Alternative Medicine, and Peer Narrative) with
two search contexts (Symptom Noticing and Active Treatment), using 30
sock-puppet accounts. GPT-5.4-mini generated 84 condition-specific queries from
\textit{r/breastcancer} seed titles, and the automated collection retrieved up
to 25 results per query, yielding 9,020 search results.
GPT-5.4-nano separately annotated comment-masked video screenshots and captured
comment text for the subsequent misinformation analyses.}
    \label{fig:data_annotation}
\vspace{-0.5cm}
\end{figure}

\textbf{Query framing.}  The primary experimental factor is the language used to express an information need. \emph{Medical Information} and \emph{Alternative Medicine} express opposing stances toward conventional care---acceptance and rejection---and together define the trust dimension around which the audit is built.
\emph{Peer Narrative}, on the other hand, takes no position on that dimension and aims to disentangle lay, non-clinical language from alternative-medicine framing. 

\emph{Medical Information} queries express acceptance of the clinical pathway and
ask how it works (\textit{``core needle biopsy breast,'' ``port care during
chemotherapy''}). \emph{Alternative Medicine} queries express the opposite
stance in two ways: by seeking non-standard remedies (\textit{``breast swelling
herbal remedies,'' ``holistic breast cancer recovery''}) and by rejecting or
avoiding clinical care without naming any remedy (\textit{``avoid biopsy breast
lump,'' ``doctors overreact to breast lumps,'' ``refused chemo breast
cancer''}). The second group matters because those queries request no
alternative-medicine content at all; they express distrust. If they still
retrieve misinformation at high rates, the pattern cannot be explained purely as
the topical relevance of alternative-medicine vocabulary.

\emph{Peer Narrative} queries seek other patients' accounts rather than facts or
treatments, spanning both support seeking (\textit{``breast cancer not alone''})
and information seeking through others' experience (\textit{``people who ignored
breast changes''}). This is neither a midpoint between the two poles nor an
uncommitted searcher: it is a distinct and equally directed orientation with its
own definite goal. Its role in the design is as a control. Because it uses lay,
non-clinical vocabulary while taking no position on conventional versus
alternative treatment, it tests whether elevated misinformation exposure follows
from lay language in general or specifically from the alternative-medicine pole
of the trust axis.

All three framings represent searchers who have already committed to an
information frame. The design therefore contains no condition for an
\emph{undirected} searcher---one with no clinical, alternative, or peer
orientation, who does not yet know what to look for. That searcher is arguably
the most policy-relevant case, since they are the most open to being moved by
whatever the ranking returns; we return to this gap in
Section~\ref{sec:limitations}.

\textbf{Search context.} The second factor represents two stages of the breast
cancer journey. \emph{Symptom Noticing} captures searches made before diagnosis,
while interpreting symptoms; \emph{Active Treatment} captures searches made after
diagnosis, while making or undergoing treatment decisions.

\textbf{Experimental controls.} All conditions share the same browser automation, collection procedures, geographic location, account settings, and simulated user profile (a 45-year-old woman, daily TikTok user, moderate health literacy). The only systematic difference across conditions is the wording of the search queries, which isolates the association between query framing and the results TikTok returns.

\textbf{Dataset.} We created 30 fresh TikTok accounts, five per condition, to capture across-account variability. Each condition was assigned 14 unique queries, giving 84 queries in total. The collection produced 420 observed search sessions and 9,020 search results, of which 7,199 were cancer-relevant. We define a \emph{search session} as one account issuing one query and collecting up to 25 top results. Collection ran May 15--21, 2026, on the TikTok web client from the United States. We focus on breast cancer as one of the most common cancers in the United States~\citep{siegel2024}, which also provides a large, active Reddit community for grounding query language.

The automated agent does not like, comment on, save, or share any posts, or follow any other accounts. However, it necessarily opened videos to capture screenshots and comment sections, so TikTok may have recorded views and clicks. We therefore interpret the results as observed exposure under a standardized search-and-view personalization rather than the output of a stateless search engine.

\vspace{-15pt}
\subsection{Query Generation}
\vspace{-10pt}

Because query framing is the manipulated variable, query construction determines what the audit measures. We generated queries by grounding an LLM in naturally occurring breast cancer discussions. We collected the 1,155 most-commented post titles from r/breastcancer using RedditExtractoR~\citep{reddit_extractor}. Ordering by comment count selects for the story types the community engages with most, so the seed vocabulary reflects prevalent phrasings. These titles were not issued as search queries and served as examples of community language.

For each of the six conditions, we prompted GPT-5.4-mini with (i) a description of the target framing and search context and (ii) the Reddit seed titles, asking it to produce short, TikTok-style queries in that framing. The full prompts can be found in Appendix~\ref{app:query_prompts}. The model was instructed to generate 14 queries per condition---two per day over the seven-day experiment---for 84 unique queries. All queries were manually reviewed for consistency with the assigned framing and search context (Appendix, Tables~\ref{tab:symptom_noticing_queries}
and~\ref{tab:active_treatment_queries}.)

\vspace{-15pt}
\subsection{Content Annotation}
\vspace{-10pt}

Over the seven-day experiment, each sock-puppet account searched TikTok twice per day using queries assigned to its experimental condition. For each search, the automated collection pipeline opened up to the top 25 returned videos and captured five screenshots spanning each video's playback. For annotation, we
selected the third screenshot from each set and used a layout-aware masking
procedure to exclude the comment panel while preserving the video and visible
post information. Using a structured GPT-5.4-nano prompt, the model classified
(i) cancer relevance; (ii) the presence of cancer misinformation; and, when
misinformation was present, (iii) its type and (iv) its topic. More than one misinformation type or topic could be assigned.
The complete prompt, output schema, and classification rules are provided in
Appendix~\ref{appendix:prompt}.

\begin{table*}[t]
\centering
\small
\caption{Possible misinformation rates among cancer-relevant search results by query framing and search context. The Alternative Medicine--Medical Information gap is 8.6$\times$ in the Symptom Noticing context and 7.6$\times$ in the Active Treatment context.}
\label{tab:rates}
\setlength{\tabcolsep}{10pt}
\begin{tabular}{llrrr}
\toprule
Query framing & Search context & Misinfo / $n$ & Rate & 95\% Wilson CI \\
\midrule
Medical Information & Symptom Noticing & 88 / 1405 & 6.3\% & [5.1\%, 7.7\%] \\
Medical Information & Active Treatment & 87 / 1229 & 7.1\% & [5.8\%, 8.7\%] \\
Peer Narrative & Symptom Noticing & 41 / 1074 & 3.8\% & [2.8\%, 5.1\%] \\
Peer Narrative & Active Treatment & 33 / 1032 & 3.2\% & [2.3\%, 4.5\%] \\
Alternative Medicine & Symptom Noticing & 547 / 1011 & 54.1\% & [51.0\%, 57.2\%] \\
Alternative Medicine & Active Treatment & 775 / 1448 & 53.5\% & [50.9\%, 56.1\%] \\
\bottomrule
\end{tabular}
\end{table*}

\begin{table*}[t]
\centering
\small
\caption{Logistic regression predicting possible misinformation among
cancer-relevant results. Medical Information framing in the Symptom Noticing context is
the reference condition. Standard errors are two-way clustered by account
and query.}
\label{tab:regression}
\begin{tabular}{lrrrr}
\toprule
Predictor & $\hat\beta$ & OR & 95\% CI & $p$ \\
\midrule
Intercept (Medical Information, Symptom Noticing) & -2.71 & 0.07 & [0.04, 0.11] & $<0.001$ \\
Alternative Medicine & +2.87 & 17.64 & [8.35, 37.29] & $<0.001$ \\
Peer Narrative & -0.52 & 0.59 & [0.31, 1.15] & 0.121 \\
Active Treatment & +0.13 & 1.14 & [0.58, 2.24] & 0.703 \\
Alternative Medicine $\times$ Active Treatment & -0.15 & 0.86 & [0.33, 2.26] & 0.754 \\
Peer Narrative $\times$ Active Treatment & -0.31 & 0.73 & [0.25, 2.09] & 0.558 \\
\bottomrule
\end{tabular}
\end{table*}

The three annotation dimensions used in this study---misinformation presence,
type, and topic---were adapted from a multidimensional taxonomy of cancer
misinformation developed in related work~\citep{ours_reddit_taxonomy}. That
framework was originally developed for Reddit posts; here, we retained the
dimensions that could be applied to video content.

\noindent\textbf{Presence of misinformation.}
Following prior health-misinformation research, we treated visible content as
possible cancer misinformation when it presented, promoted, or endorsed a
specific false, misleading, or scientifically unsupported cancer claim that
could cause harm~\citep{chou2018addressing,loeb2024}. Merely mentioning,
questioning, criticizing, or correcting a false claim did not receive a
positive label. Personal experience alone was also insufficient unless it
promoted a harmful unsupported claim or encouraged rejection or avoidance of
standard cancer care. 

The prompt provided five illustrative claim families to guide this binary
decision: unproven or ineffective treatments; misrepresentation or selective
use of facts; outdated or decontextualized medical information; conspiracy
narratives; and commercial promotion based on unsupported cancer claims
\citep{loeb2024}. These examples guided misinformation detection but were not
the categories assigned to positive videos.

\noindent\textbf{Misinformation type.}
For positive cases, misinformation type describes \emph{how} the misleading
claim is constructed. Following the framework developed in related
work~\citep{ours_reddit_taxonomy}, we adapted seven mechanisms from established
information-disorder typologies: fabricated content, misleading content, false
context, false connection, imposter content, manipulated content, and
satire/parody~\citep{wardle2017information}. An \emph{Other} option captured
claims that did not fit these seven mechanisms, giving eight possible output
labels.

\noindent\textbf{Misinformation topic.}
Misinformation topic describes \emph{what} the misleading claim concerns. We
operationalized seven cancer-specific topics: causation, risk factors, and
prevention; diagnosis and screening; safety and effectiveness of conventional
treatments; unproven or alternative treatments; conspiracy and institutional
distrust; morality, religion, and ideology; and medical autonomy and medical
freedom. An \emph{Other} option captured remaining topics. These categories are
our operationalization of prior health- and cancer-misinformation research,
rather than a taxonomy reproduced verbatim from any single source
\citep{kata2010postmodern,loeb2024}.

For every positive misinformation label, the classifier was required to assign
exactly one primary type and one primary topic. It could assign up to two
additional types and two additional topics when each was independently
supported by visible evidence. The classifier also returned a short neutral
paraphrase of the claim and one to three pieces of visible evidence. The topic
and type distributions combine primary and additional labels; consequently,
the categories are not mutually exclusive.

\noindent\textbf{Comment annotation.}
Captured comments were annotated in a separate GPT-5.4-nano text-classification
pass. Comment misinformation presence, type, and topic were determined only
from the captured comment strings, using the same fixed type and topic
vocabularies. The creator caption and on-screen text were supplied only to
assess whether comments supported or challenged the video's visible message.
The comment classifier did not receive the video screenshot or its
misinformation label. The comment-classification rules are summarized in
Appendix~\ref{app:comment-prompt}.

\vspace{-20pt}

\subsection{Statistical Analysis}
\vspace{-10pt}

\textbf{Exposure rates.} Misinformation rates are reported with 95\% Wilson confidence intervals throughout the results section. To assess whether misinformation is concentrated near the top of the result list, we compute \emph{position-weighted exposure}, a weighted average in which each occurrence is weighted by $1/\mathrm{rank}$:
\begin{equation}
\mathrm{PW} = \frac{\sum_i m_i / \mathrm{rank}_i}{\sum_i 1 / \mathrm{rank}_i},
\label{eq:posweight}
\end{equation}
where $m_i = 1$ for a misinformation-labeled occurrence $i$ and $0$ otherwise, and $\mathrm{rank}_i$ is its position in the result list. Normalizing by the total weight keeps PW on the same scale as the raw rate, so a PW close to the raw rate indicates a roughly uniform rank profile, while a PW well above it indicates
top-heavy concentration.


\textbf{Video-level regression.}
The primary inferential model is a binomial logistic regression predicting
possible misinformation among cancer-relevant search results. Query
framing, search context, and their interaction are included as fixed effects,
with Medical Information in the Symptom Noticing context as the reference
condition. We calculate two-way cluster-robust standard errors by account and
query to account for dependence among results returned to the same account and
results generated by repeated uses of the same query. A joint Wald test
evaluates the two query-framing--search-context interaction coefficients.

We evaluate robustness using three additional specifications: a deduplicated
analysis retaining the highest-coverage collection for each account,
condition, and day; generalized estimating equations grouped separately by
account and query; and a Bayesian variational logistic mixed model with crossed
random intercepts for account and query. The mixed-model results are reported
in Appendix~\ref{app:glmm}.

\textbf{Taxonomy and comment analyses.}
Misinformation type and topic distributions are calculated among
cancer-relevant exposures labeled as containing possible misinformation. We
count each assigned category at most once per exposure and include both primary
and additional labels, so percentages may sum to more than 100\%. Associations
between the video classification and comment outcomes are summarized using
odds ratios and Fisher's exact tests.

\vspace{-12pt}
\section{Results}
\vspace{-10pt}
\label{sec:results}

\subsection{Misinformation Rates.} 
\vspace{-10pt}

Of the 7,199 cancer-relevant search results, 1,571 (\textbf{21.8\%}) were labeled as containing possible misinformation. Across all 9,020 retrieved videos, including results that were not cancer-relevant, the corresponding rate was 17.4\%. Because a result that is not cancer-relevant cannot contain cancer misinformation under our annotation definition, we use the cancer-relevant set as the primary denominator unless otherwise noted. Table~\ref{tab:rates} reports rates by framing and context with Wilson confidence intervals.
Cancer relevance itself varies by condition. Alternative Medicine queries under Symptom Noticing produced the smallest cancer-relevant pool ($n = 1{,}011$), because queries such as \textit{``holistic breast health tips''} and
\textit{``breast lump massage''} retrieve substantial general-wellness content with no explicit cancer framing, which our relevance criterion excludes. Alternative Medicine framing therefore retrieves a pool that is both topically looser and, as we show next, far more misinformation-dense.

\begin{figure*}[t]
\centering
\includegraphics[width=0.85\linewidth]{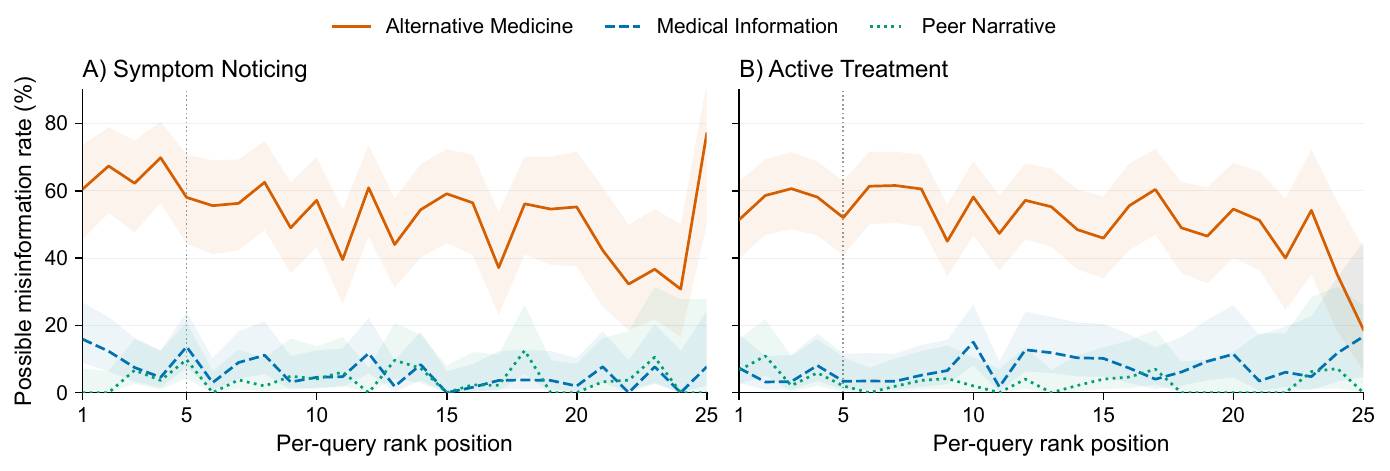}
\vspace{-10pt}
\caption{Possible misinformation rates by per-query rank position.
Panel~A: Symptom Noticing; Panel~B: Active Treatment.
Each panel includes the Alternative Medicine, Medical Information, and
Peer Narrative query framings. Shaded bands show 95\% Wilson confidence
intervals for the rate at each rank; the vertical dotted line marks rank~5.}
\vspace{-10pt}
\label{fig:rank}
\end{figure*}

\vspace{-15pt}
\subsection{Framing Differences and \\ Their Stability Across Contexts}
\vspace{-10pt}
\label{sec:res-framing}

Among cancer-relevant results (Table~\ref{tab:rates}), Alternative Medicine
queries returned possible misinformation at \textbf{8.6 times} the Medical
Information rate in the Symptom Noticing context (54.1\% vs.\ 6.3\%) and
\textbf{7.6 times} in the Active Treatment context (53.5\% vs.\ 7.1\%).
Peer Narrative queries returned the lowest observed rates, 3.8\% and 3.2\%,
respectively.

The regression (Table~\ref{tab:regression}) confirms that Alternative Medicine is the only non-intercept coefficient that reaches significance.
Relative to Medical Information under Symptom Noticing, Alternative Medicine
framing raises the odds of a possible-misinformation label
(OR\,=\,17.64, 95\% CI [8.35, 37.29], $p<0.001$). The Peer
Narrative--Medical Information contrast does not reach the 
threshold (OR\,=\,0.59, 95\% CI [0.31, 1.15], $p=0.121$), and neither does the
Active Treatment--Symptom Noticing contrast (OR\,=\,1.14, 95\% CI
[0.58, 2.24], $p=0.703$). Neither interaction coefficient is significant, and
the joint interaction test is also not significant
(Wald $\chi^2(2)=0.35$, $p=0.841$). Standard errors are clustered two-way by
account and query; because five replicate accounts issued the same query,
results within a query are not independent, and treating them as such would
understate uncertainty.

The results suggest that exposure is governed by query framing and not detectably by the stage of the patient journey. 
Neither the Active Treatment main effect
($\mathrm{OR}=1.14$, 95\% CI $[0.58, 2.24]$, $p=0.703$) nor the
framing-by-context interaction (joint Wald $\chi^2(2)=0.35$, $p=0.841$)
reached significance: the framing gap is as
large for a woman interpreting a new lump as for one weighing a treatment
decision, within every framing. In effect, the search system returns the
same misinformation prevalence to a user issuing symptom-stage queries as
to one issuing active-treatment queries, even though the latter are tied
to consequential and often irreversible treatment decisions.

Moreover, Peer Narrative provides the control the design was built for. It uses informal,
lay language (\textit{``breast cancer support community,'' ``women sharing chemo
stories''}) with no clinical or alternative-medicine vocabulary, and it shows
\emph{no elevation} in exposure---the point estimate is below the Medical Information baseline,
though not significantly. Elevated misinformation is therefore not a general
property of lay or non-clinical phrasing; it tracks the alternative-medicine
orientation specifically. The conditions differ in topic and intent as well as
wording, so matched-paraphrase designs would be needed to separate those
components.

The most extreme was \textit{``natural ways to reduce breast lump,''} a
Symptom Noticing query, for which 87 of 94 cancer-relevant results
(\textbf{92.6\%}) were labeled as containing possible misinformation.

\begin{figure*}[t]
\centering
\includegraphics[width=0.85\textwidth]{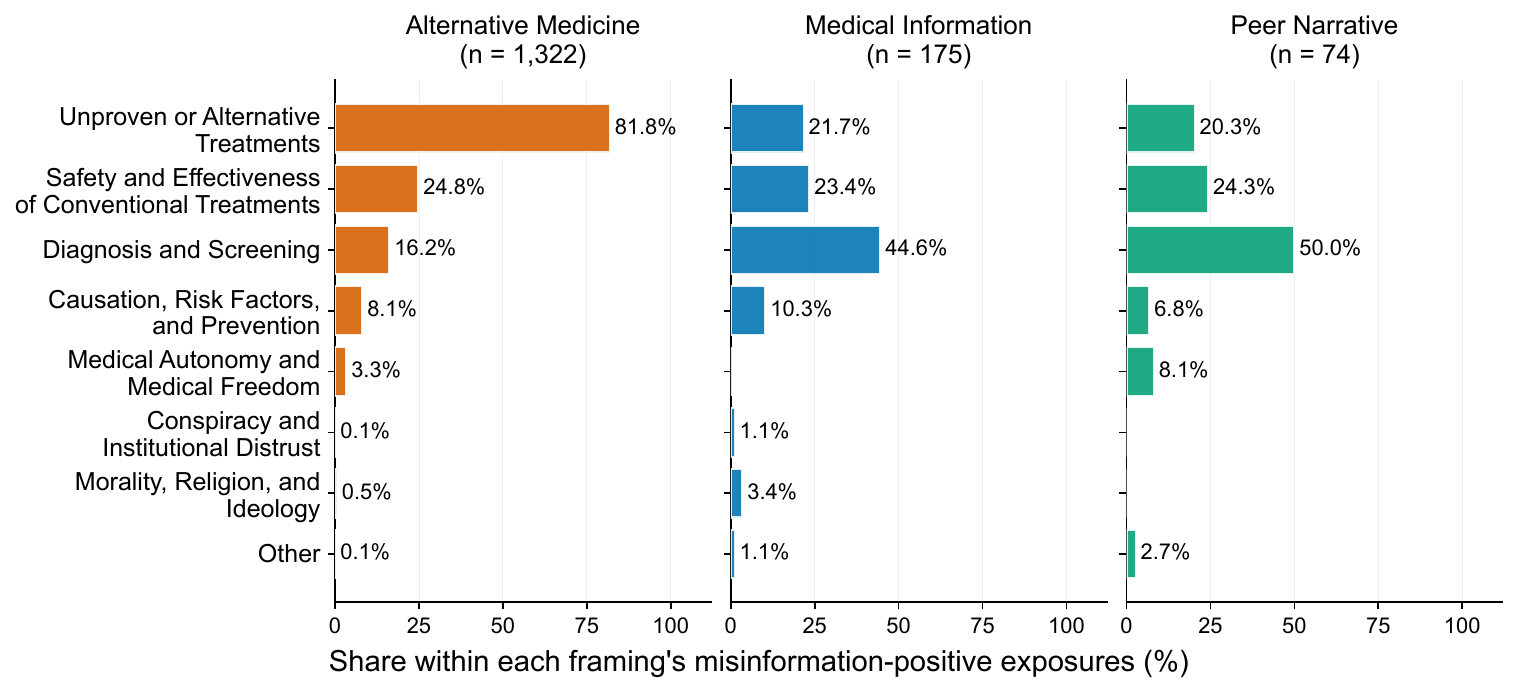}
\vspace{-8pt}
\caption{Distribution of misinformation topics by query framing among
cancer-relevant search results classified as containing possible
misinformation, pooling the Symptom Noticing and Active Treatment contexts.
Each panel uses the misinformation-positive exposures within that framing as
its denominator (Alternative Medicine, $n=1{,}322$; Medical Information,
$n=175$; Peer Narrative, $n=74$). Bars include all assigned labels. Because an
exposure could receive more than one topic, percentages within a panel may sum
to more than 100\%.}
\vspace{-8pt}
\label{fig:misinformation-topics}
\end{figure*}

\begin{figure}[t]
\centering
\includegraphics[width=\linewidth]{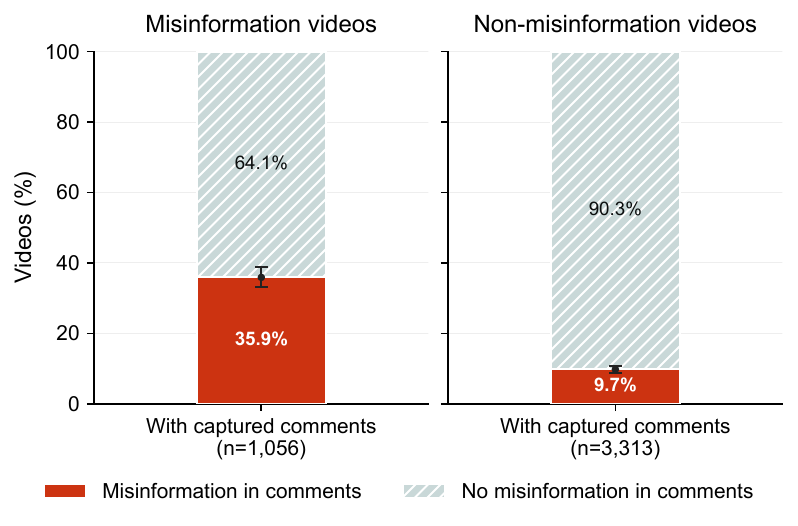}
\vspace{-8pt}
\caption{Misinformation in captured comments for misinformation and
non-misinformation videos. Each panel includes only videos with captured
comment text. Stacked bars show the percentages whose captured comments
contained and did not contain misinformation. Black points and error bars mark
the comment-misinformation percentage and its 95\% Wilson confidence interval.
Video labels were assigned from screenshots with the comments removed.}
\vspace{-5pt}
\label{fig:comments}
\end{figure}

\vspace{-10pt}
\subsection{Rank Profile of Exposure}
\vspace{-10pt}
\label{sec:rank}

For Alternative Medicine searches in the Symptom Noticing context
(Figure~\ref{fig:rank}A), possible misinformation appears throughout the
ranked list, although it is somewhat more concentrated near the top. The pooled
rate is \textbf{63.8\%} across positions 1--5 and \textbf{48.9\%} across
positions 11--25. Inverse-rank-weighted exposure is \textbf{59.2\%}, compared
with the raw cancer-relevant rate of \textbf{54.1\%}, indicating some
top-weighting but not confinement to the first few positions. Among
search-result lists containing at least one misinformation-labeled result, the
median first misinformation rank is \textbf{2.0}. Overall, 68 of 76
Alternative Medicine search-result lists (\textbf{89.5\%}) in the Symptom
Noticing context contain at least one misinformation-labeled result within the
first ten positions.

\vspace{-10pt}
\subsection{Intent--Exposure Mismatch}
\vspace{-10pt}

Because Alternative Medicine queries are themselves oriented toward
alternative-medicine content, the framing gap reflects both retrieval behavior and
the topical relevance of the queries. Different query vocabularies are associated
with content pools that have very different misinformation prevalence. The Medical Information framing provides a useful comparison: even with explicitly clinical
vocabulary, possible misinformation appeared in 6.3\%--7.1\% of cancer-relevant results.
The query \textit{``evidence-based breast cancer recovery''} returned possible
misinformation in 15 of 55 cancer-relevant results (27.3\%), showing that substantial misinformation can appear even for a query with clearly clinical intent. This is consistent with prior search-audit evidence that search systems can surface misinformation despite explicit query intent.

\vspace{-10pt}
\subsection{Topics of Possible Misinformation}
\vspace{-10pt}
\label{sec:results-topics}

We also examined the topics assigned to cancer-relevant search-result
exposures classified as containing possible misinformation. Across all query
framings, unproven or alternative treatments were the most common topic
(1,135 of 1,571; 72.2\%), followed by the safety and effectiveness of
conventional treatments (387; 24.6\%) and diagnosis and screening
(329; 20.9\%).

The topic distribution differed descriptively across query framings
(Figure~\ref{fig:misinformation-topics}). Among Alternative Medicine
misinformation-positive exposures, 81.8\% involved unproven or alternative
treatments, 24.8\% involved the safety or effectiveness of conventional
treatments, and 16.2\% involved diagnosis or screening. In contrast, diagnosis
and screening was the most common topic among misinformation-positive Medical
Information (44.6\%) and Peer Narrative exposures (50.0\%). Unproven or
alternative treatments accounted for 21.7\% and 20.3\% of the corresponding
exposures, respectively. These comparisons are descriptive because the query
framings differ in topic and intent and contain substantially different numbers
of misinformation-positive exposures.

\vspace{-10pt}
\subsection{Comment-Section Concordance}
\vspace{-10pt}

At the time of data collection, comments were \emph{visible}
for 1,224 of 1,571 videos labeled as containing possible misinformation (78.0\%), compared with 4,019 of 5,586 non-misinformation videos (71.9\%; odds
ratio\,=\,1.38, $p<0.001$). Among those videos with visible comments, comment text was successfully
captured and annotated for 1,056 misinformation-labeled videos and 3,313
non-misinformation videos.
Within that subset, misinformation appeared in the comments of 379 of 1,056 misinformation-labeled videos (\textbf{35.9\%}), compared with 323 of 3,313 non-misinformation videos (\textbf{9.7\%}; odds ratio\,=\,5.18,
$p<0.001$; Figure~\ref{fig:comments}). Video labels and comment labels were
generated in separate annotation passes: video labels used screenshots with
the comments removed, whereas comment labels used only the captured comment
text. We therefore report this association as co-occurrence, not as evidence
that comments cause or affect misinformation exposure.

\begin{table*}[t]
\centering
\small
\caption{Performance and pairwise agreement for binary misinformation
classification. Nano = GPT-5.4-nano; Mini = GPT-5.4-mini; Gemini =
Gemini 3.1 Pro Preview. For model comparisons, all models received the
identical prompt and comment-excluded screenshots.}
\label{tab:classifier}

\begin{minipage}{0.82\textwidth}
\centering
\textbf{(a) Performance against human labels}

\vspace{3pt}
\begin{tabular*}{\linewidth}
{@{\extracolsep{\fill}}lrrrrrrr@{}}
\toprule
\textbf{Model} & \textbf{$n$} & \textbf{P} & \textbf{R} &
\textbf{$F_1$} & \textbf{Acc} & \textbf{MCC} & \textbf{$\kappa$} \\
\midrule
Nano
  & 50 & 0.84 & \textbf{0.84} & \textbf{0.84}
  & \textbf{0.88} & 0.75 & \textbf{0.75} \\
Mini
  & 50 & \textbf{1.00} & 0.68 & 0.81
  & \textbf{0.88} & \textbf{0.76} & 0.73 \\
Gemini
  & 50 & 0.92 & 0.63 & 0.75
  & 0.84 & 0.66 & 0.64 \\
\bottomrule
\end{tabular*}
\end{minipage}

\vspace{10pt}

\begin{minipage}{0.62\textwidth}
\centering
\textbf{(b) Pairwise model agreement}

\vspace{3pt}
\begin{tabular*}{\linewidth}
{@{\extracolsep{\fill}}lrrrr@{}}
\toprule
\textbf{Model pair} & \textbf{$n$} &
\textbf{Agreement} & \textbf{MCC} & \textbf{$\kappa$} \\
\midrule
Nano--Mini
  & 386 & 0.92 & 0.74 & 0.73 \\
Nano--Gemini
  & 386 & 0.92 & 0.74 & 0.73 \\
Mini--Gemini
  & 386 & \textbf{0.96} & \textbf{0.87} & \textbf{0.87} \\
\bottomrule
\end{tabular*}
\end{minipage}

\vspace{4pt}

\begin{minipage}{0.92\textwidth}
\footnotesize
\textit{Note.} Precision, recall, and $F_1$ treat the human label as the
reference and misinformation as the positive class. Model--model comparisons
use symmetric agreement measures because neither model is treated as ground
truth. Panel~(b) reports pairwise agreement on the 386-item evaluation sample.
All metrics evaluate binary misinformation presence, not misinformation type
or topic. Bold values denote the largest value within each panel; ties are
retained.
\end{minipage}
\end{table*}
\vspace{-10pt}
\subsection{Classifier Validation}
\label{sec:res-validation}
\vspace{-10pt}

On the 50-item human-labeled subset, which contained 19 positive and 31
negative cases, GPT-5.4-nano produced 16 true positives, 28 true negatives,
3 false positives, and 3 false negatives. Its precision, recall, and $F_1$
were each 0.84, with an accuracy of 0.88 and Cohen's $\kappa=0.75$.
Errors were evenly divided between false positives and false negatives.

To assess whether the binary labels were specific to GPT-5.4-nano, we applied
the identical prompt, output schema, and comment-excluded screenshots to two
comparison models: GPT-5.4-mini and Gemini 3.1 Pro Preview. Pairwise agreement
was evaluated on a 386-item evaluation sample. Nano agreed with Mini and Gemini
on 91.7\% of cases in both comparisons ($\kappa=0.73$ for each), while Mini
and Gemini agreed on 96.4\% of cases
($\kappa=0.87$; Table~\ref{tab:classifier}).

GPT-5.4-nano achieved the highest $F_1$ and $\kappa$ against the human labels,
while GPT-5.4-mini achieved perfect precision and the highest MCC. Mini and
Gemini were more conservative: their errors against the human labels were
primarily false negatives. The strong Mini--Gemini agreement, together with
each model's agreement with Nano, indicates that the binary signal is not
unique to one model, although the models differ in where they place the
decision boundary.

Finally, the 50-item human-labeled subset is too small to estimate reliable
framing- or context-specific error rates. The human labels evaluate only
misinformation presence; the misinformation-type and misinformation-topic
assignments were not independently human-validated.
Additional condition-level visualizations, per-query and account-specific
variation, and misinformation-type distributions are reported in
Appendix~\ref{app:supp-results}.

\vspace{-10pt}
\section{Discussion}
\vspace{-10pt}
\label{sec:discussion}

Our findings show two descriptive patterns in TikTok search results. First, different query framings are associated with returned content pools that
have sharply different misinformation prevalence: Alternative Medicine queries returned possible misinformation in 54.1\% of cancer-relevant results in Symptom Noticing and 53.5\% in Active Treatment. Second, clinically framed queries did not eliminate misinformation: Medical Information queries still returned possible misinformation in 6.3\%--7.1\% of cancer-relevant results despite using clinical search terms. The second pattern extends \citet{hussein2020}'s YouTube evidence that search systems can surface misinformation under health-related queries to a mobile-first, short-video setting with high-stakes cancer information.

In this audit, exposure differs across query framings while the simulated demographics and health-literacy profile are held fixed;
Alternative Medicine queries have higher misinformation rates, whereas Peer Narrative queries show no elevation relative to Medical Information. Platform audits should therefore
measure differential exposure across query framings without attributing the
pattern to patient literacy or identity.

A natural expectation is
that a responsible ranking system would apply \textit{greater} scrutiny to
queries tied to active treatment decisions, where misinformation is most
consequential: content that discourages chemotherapy, radiation, or
surgery can translate directly into treatment delay, reduction, or
abandonment, and these decisions are time-sensitive and frequently
irreversible. Our data show the opposite of such stage-sensitivity---not
that the system protects treatment-stage searchers less, but that it does
not distinguish between the stages at all. Exposure is governed almost
entirely by query framing and is statistically flat across the
Symptom Noticing and Active Treatment contexts. We interpret this
indifference itself as a safety gap: the platform treats a user weighing
whether to begin chemotherapy no differently from one interpreting an
ambiguous symptom, despite the sharply higher stakes of the former. This
is an inference about how platforms \textit{should} weight query context,
not a causal claim about the ranking algorithm; our audit measures
exposure, not intent.

Three intervention points follow from the data:
(1)~\textit{Reducing misinformation in the retrievable pool}: Because misinformation remained common beyond the highest-ranked results for Alternative Medicine queries, changes limited to the first few positions may
not substantially reduce exposure. A more effective intervention may require reducing the prevalence of misinformation in the pool of content retrieved for these queries, alongside query-level safeguards such as authoritative redirection.
(2)~\textit{Comment-section interventions}: among videos with captured comment
text, comments contained misinformation for 35.9\% of misinformation videos,
compared with 9.7\% of non-misinformation videos. Corrective nudges,
authoritative pinned replies, and links to verified health information may
therefore be useful to evaluate.
(3) \textit{Stage-sensitive safeguards}: because misinformation exposure
does not decline for treatment-decision queries, and because the
highest-risk framing remains saturated in this context---Alternative
Medicine queries returned possible misinformation in 53.5\% of
cancer-relevant results during Active Treatment, $7.6\times$ the Medical
Information rate---authoritative redirection and demotion may warrant
stronger application to queries expressing treatment-refusal or
treatment-substitution intent (e.g., ``refused chemo breast cancer,''
``breast cancer without radiation,'' ``chemo alternatives breast
cancer''). The absence of stage-sensitivity in current results suggests
such targeting is not presently occurring.

Our use of LLM-generated, condition-specific queries within controlled sock-puppet agents illustrates a scalable, reproducible approach to controlled query-vocabulary auditing: grounding query generation in cancer-community Reddit titles provides examples of patient language and concerns while LLM generation ensures platform-appropriate phrasing, generalizing to other health topics and platforms.


GPT-5.4-nano reached $\kappa=0.75$ against the 50-item human-labeled subset
and produced equal numbers of false positives and false negatives. On the
386-item model-comparison sample, its agreement with GPT-5.4-mini and Gemini
3.1 Pro Preview was $\kappa=0.73$ in both cases, while Mini and Gemini reached
$\kappa=0.87$. These results support model-assisted annotation for aggregate
auditing and human-in-the-loop triage, not autonomous content moderation.

Misleading Content was the dominant misinformation type under all three query
framings. Topic distributions differed more clearly: 81.8\% of
misinformation-positive Alternative Medicine results concerned unproven or
alternative treatments, whereas Diagnosis and Screening was the most common
topic under Medical Information (44.6\%) and Peer Narrative (50.0\%).

\vspace{-10pt}
\section{Limitations}
\vspace{-10pt}
\label{sec:limitations}

The misinformation labels are model-assisted audit labels, not definitive
clinical judgments. GPT-5.4-nano reached $\kappa=0.75$ against a 50-item
human-labeled subset, but this subset is too small to estimate reliable
framing- or context-specific error rates. The human labels evaluate only
misinformation presence; the misinformation-type and misinformation-topic
assignments were not independently human-validated.

The VLM evaluated the third screenshot from each five-screenshot sequence.
Claims visible only earlier or later in a video may therefore have been missed.

This audit covers TikTok search only, not the For-You-Page feed or combined search-plus-feed exposure. Account history may also accumulate during the seven-day protocol, so results describe observed exposure under our framing-conditioned search-and-view design.
Query conditions differ in topic and intent as well as wording. Alternative Medicine queries are designed to seek alternative-medicine content, so the framing gap partly reflects the content pool those queries retrieve.
LLM-generated queries may also miss some real patient or caregiver vocabulary, and the one-week collection window cannot capture temporal variation.

\vspace{-10pt}
\section{Conclusion}
\vspace{-10pt}
\label{sec:conclusion}

In a 30-account audit of 9,020 TikTok search results, 1,571 of
the 7,199 cancer-relevant results (21.8\%) were labeled as containing possible
misinformation. Alternative Medicine queries returned 8.6 times the Medical
Information rate in the Symptom Noticing context and 7.6 times the rate in the
Active Treatment context. Medical Information queries nevertheless returned
misinformation rates of 6.3\%--7.1\%, while Peer Narrative queries produced
the lowest observed rates. For Alternative Medicine searches, misinformation
remained common beyond the highest-ranked results in both search contexts.
A separate classification of captured comment text also showed that
misinformation in comments occurred more frequently on videos labeled as
misinformation, although this association is descriptive. Together, these
findings show that query framing is strongly associated with misinformation
exposure in TikTok cancer searches. The LLM/VLM pipeline supports scalable
human-in-the-loop auditing and triage, not autonomous moderation.

\bibliography{References}

\appendix
\counterwithin{table}{section}
\renewcommand{\thetable}{\thesection.\arabic{table}}

\makeatletter
\@addtoreset{figure}{section}
\makeatother
\renewcommand{\thefigure}{\thesection.\arabic{figure}}

\section{Query-Generation Prompt}
\label{app:query_prompts}

The same prompt structure was used across all six framing--context conditions,
with bracketed fields replaced by condition-specific content.

\begin{table*}[t]
\caption{Shared prompt template used to generate the TikTok search queries.
Bracketed fields were replaced with the relevant search context,
query-framing instructions, and Reddit titles.}
\label{tab:query-prompt-template}

\begin{tcolorbox}[
  auditbox,
  colback=white,
  colframe=black!25,
  boxrule=0.4pt,
  arc=0.8mm,
  left=2mm,
  right=2mm,
  top=1.5mm,
  bottom=1.5mm
]
\footnotesize
\raggedright

\textbf{System message}\par
You are a precise JSON-only responder. Return only valid JSON arrays
with no extra text.

\tcbline

\textbf{User prompt}\par
You are helping a research team study cancer misinformation on TikTok.

\medskip

\begin{tabular}{@{}ll@{}}
\textbf{Cancer type:}   & breast cancer \\
\textbf{Patient phase:} & \emph{[search context]} \\
\textbf{Persona:}       & \emph{[query framing]} \\
\end{tabular}

\medskip

\textbf{Persona behavior}\par
\emph{[condition-specific instructions from
Table~\ref{tab:query-condition-instructions}]}

\medskip

Below are real Reddit post titles from breast cancer patients at this
phase. Use these only to understand authentic patient language and
concerns. Do not copy these titles---generate original TikTok search
queries:

\smallskip

\emph{[phase-specific Reddit titles]}

\tcbline

\noindent
\begin{minipage}[t]{0.64\linewidth}
\raggedright

\textbf{Task}\par
Generate exactly 14 TikTok search queries this persona would type across
7 days. Assign exactly 2 queries per day (days 1 through 7).

\medskip

\textbf{Rules}
\begin{itemize}[leftmargin=1.3em,itemsep=1pt,topsep=3pt]
    \item Queries must strictly match this persona's intent---no exceptions.
    \item NO escalation or change across days---same intent throughout.
    \item Vary the topic across days but NEVER vary the stance.
    \item 3 to 7 words per query---natural TikTok search style.
    \item Breast cancer specific---not generic cancer queries.
    \item No Reddit-style fragments---clean search queries only.
    \item All 14 queries must be meaningfully different from each other.
\end{itemize}

\end{minipage}
\hfill
\begin{minipage}[t]{0.32\linewidth}
\raggedright

\textbf{Required output}\par
Return only a valid JSON array of exactly 14 objects.

\medskip

Each object must contain:

\begin{itemize}[leftmargin=1.3em,itemsep=2pt,topsep=3pt]
    \item \texttt{"day"}: an integer from 1 to 7, with exactly 2 queries per day;
    \item \texttt{"query"}: the search-query text.
\end{itemize}

No explanation. No Markdown. Only the JSON array.

\end{minipage}

\end{tcolorbox}
\end{table*}

\begin{table*}[p]
\centering
\caption{Condition-specific instructions inserted into the query-generation
prompt.}
\label{tab:query-condition-instructions}
\footnotesize
\renewcommand{\arraystretch}{1.15}

\begin{ruledtabular}
\begin{tabular}{@{}p{0.16\textwidth}p{0.16\textwidth}p{0.65\textwidth}@{}}
\textbf{Framing} & \textbf{Context} & \textbf{Instructions} \\
\hline

Medical Information &
Symptom Noticing &
The searcher has noticed breast symptoms and immediately seeks clinical
information. Queries ask what symptoms mean, when to see a doctor, what tests
to expect, and how procedures such as mammography, MRI, and biopsy work.
Queries must not seek personal stories, alternative explanations, or ways to
avoid medical care. \\

Alternative Medicine &
Symptom Noticing &
The searcher is skeptical of conventional medicine and seeks natural
explanations, remedies, holistic approaches, and ways to address breast
symptoms without doctors. Queries must not request clinical procedures,
medical information, or emotional-support stories. \\

Peer Narrative &
Symptom Noticing &
The searcher seeks personal stories, reassurance, shared experiences, and
connection with others who noticed similar breast changes. Queries should
reflect varied emotions, including confusion, denial, reassurance seeking,
feeling overwhelmed, and occasional fear. Queries must not request medical
advice, alternative treatments, or treatment opinions, and must not imply an
existing diagnosis. \\

Medical Information &
Active Treatment &
The searcher is undergoing breast-cancer treatment and seeks clinical
information about side effects, procedures, medical terminology, recovery,
and evidence-based guidelines. Queries must not seek personal stories,
alternative treatments, or reasons to question the treatment plan. \\

Alternative Medicine &
Active Treatment &
The searcher seeks natural or alternative cancer treatments, stories about
refusing or reducing conventional treatment, and content questioning
chemotherapy, radiation, or hormone therapy. Queries must not request clinical
information, emotional-support stories, or content supporting conventional
treatment. \\

Peer Narrative &
Active Treatment &
The searcher seeks treatment stories, emotional support, coping experiences,
and connection with others undergoing chemotherapy or radiation. Queries must
not request clinical information, treatment alternatives, or content
questioning the treatment plan. \\

\end{tabular}
\end{ruledtabular}
\end{table*}


\section{VLM Classification}
\label{appendix:prompt}

The production video classifier used GPT-5.4-nano
(\texttt{gpt-5.4-nano-2026-03-17}). One comment-masked third screenshot was
used for each search-result record. The model received only information visible
in that frame and was not given comments, engagement counts, the search query,
or the experimental condition. Tables~\ref{tab:vlm-schema},
\ref{tab:misinfo-types}, and~\ref{tab:misinfo-topics} summarize the structured
output and fixed classification vocabularies.

\begin{table*}[t]
\centering
\footnotesize
\caption{Structured output used for video-only VLM classification.}
\label{tab:vlm-schema}
\renewcommand{\arraystretch}{1.08}
\begin{ruledtabular}
\begin{tabular}{@{}p{0.28\linewidth}p{0.67\linewidth}@{}}
\textbf{Field} & \textbf{Definition and output rule} \\
\hline
\texttt{screenshot\_usable},
\texttt{unusable\_reason} &
Whether one opened and classifiable TikTok post was visible; otherwise, the
reason the screenshot could not be classified \\

\texttt{possible\_cancer\_relevance},
\texttt{cancer\_confidence} &
Whether the visible content concerned cancer risk, prevention, symptoms,
screening, diagnosis, treatment, survivorship, experience, or alternative
cancer therapy, with confidence from 0 to 1 \\

\texttt{misinformation\_presence} &
True only when the visible frame presented, promoted, or endorsed a specific
false, misleading, or scientifically unsupported cancer claim that could cause
harm \\

\texttt{misinformation\_type\_primary},
\texttt{misinformation\_type\_additional} &
One required primary type for a positive case and up to two independently
supported additional types \\

\texttt{misinformation\_topic\_primary},
\texttt{misinformation\_topic\_additional} &
One required primary topic for a positive case and up to two independently
supported additional topics \\

\texttt{misinformation\_claim},
\texttt{visible\_evidence} &
A short neutral paraphrase of the claim and one to three visible supporting
elements for a positive case \\

\texttt{classification\_confidence} &
Confidence in the complete annotation, ranging from 0 to 1 \\
\end{tabular}
\end{ruledtabular}
\end{table*}

\begin{table*}[t]
\centering
\footnotesize
\caption{Misinformation types used by the video classifier. Type describes
how the misinformation is constructed.}
\label{tab:misinfo-types}
\renewcommand{\arraystretch}{1.08}
\begin{ruledtabular}
\begin{tabular}{@{}p{0.25\linewidth}p{0.70\linewidth}@{}}
\textbf{Type} & \textbf{Operational definition} \\
\hline
Fabricated Content &
Completely false content with no factual basis \\

Misleading Content &
Content that distorts, exaggerates, oversimplifies, or selectively presents
facts \\

False Context &
Genuine information presented in an incorrect, inappropriate, or misleading
context \\

False Connection &
A visible headline, claim, or conclusion conflicts with its visible supporting
content; missing evidence alone does not qualify \\

Imposter Content &
Visible evidence that a person, account, organization, or source is fake,
impersonated, or falsely attributed \\

Manipulated Content &
Altered imagery, video, audio, statistics, or data used to mislead \\

Satire/Parody &
Humorous or satirical content presented so that it could reasonably be
mistaken for factual cancer information \\

Other &
Visible cancer misinformation whose construction does not fit the named types \\
\end{tabular}
\end{ruledtabular}
\end{table*}

\begin{table*}[t]
\centering
\footnotesize
\caption{Misinformation topics used by the video classifier. Topic describes
what the misinformation concerns.}
\label{tab:misinfo-topics}
\renewcommand{\arraystretch}{1.08}
\begin{ruledtabular}
\begin{tabular}{@{}p{0.31\linewidth}p{0.64\linewidth}@{}}
\textbf{Topic} & \textbf{Operational definition} \\
\hline
Causation, Risk Factors, and Prevention &
Unsupported cancer causes, distorted risk factors, or unsupported prevention
claims \\

Diagnosis and Screening &
Misleading claims about symptoms, self-diagnosis, screening, diagnostic tests,
or test safety, accuracy, or necessity \\

Safety and Effectiveness of Conventional Treatments &
Misleading claims that chemotherapy, radiation, surgery, immunotherapy, or
clinician-guided treatment is ineffective, harmful, toxic, unnecessary, or
deceptive \\

Unproven or Alternative Treatments &
Herbs, supplements, diets, fasting, detoxes, or other unproven approaches
presented as cancer treatments, cures, or substitutes for standard care \\

Conspiracy and Institutional Distrust &
Hidden cures, suppressed research, cover-ups, or unsupported claims of
deliberate deception by clinicians or institutions \\

Morality, Religion, and Ideology &
Religious, spiritual, karmic, moral, or ideological explanations of cancer or
objections to cancer care \\

Medical Autonomy and Medical Freedom &
Rights-based claims about refusing cancer care, bodily autonomy, parental
authority, or medical mandates \\

Other &
Visible cancer misinformation whose subject does not fit the named topics \\
\end{tabular}
\end{ruledtabular}
\end{table*}

For a positive misinformation label, the schema required one primary type, one
primary topic, a claim paraphrase, and visible evidence. Up to two additional
types and topics were permitted only when independently supported. The
\emph{Other} label could not be combined with a named label. For a negative
label, all type, topic, claim, and evidence fields were empty. A non-cancer
screenshot could not receive a positive cancer-misinformation label.

\subsection{Comment-Text Classification}
\label{app:comment-prompt}

Comments were classified in a separate text-only pass using the same
misinformation-presence definition and fixed type and topic vocabularies. The
classifier returned comment misinformation presence, up to three supported
types and topics, short claim paraphrases, evidence comments, and whether the
comments supported, challenged, mixed with, or were unrelated to the video's
visible message. The video screenshot and video misinformation label were not
provided to this classifier.


\section{Search Queries}
\label{app:search_queries}
The complete set of 84 TikTok search queries used in the audit is reported in
Tables~\ref{tab:symptom_noticing_queries} and~\ref{tab:active_treatment_queries}.
Queries are organized by search context and query framing, with 14 unique queries per
condition---two queries per day over the seven-day collection period. The first
table presents Symptom Noticing, and the second presents Active Treatment.

\newcommand{\querypersonarow}[1]{%
  \hline
  \multicolumn{3}{c}{%
    \rule[-0.7ex]{0pt}{3ex}\textbf{#1}%
  }\\
  \hline
}

\begin{table*}[t]
\caption{TikTok search queries used in the Symptom Noticing context.}
\label{tab:symptom_noticing_queries}
\footnotesize
\renewcommand{\arraystretch}{1.12}
\setlength{\tabcolsep}{6pt}
\begin{ruledtabular}
\begin{tabular}{
    c
    p{0.39\textwidth}
    p{0.39\textwidth}
}
\textbf{Day}
& \multicolumn{1}{c}{\textbf{Query 1}}
& \multicolumn{1}{c}{\textbf{Query 2}} \\

\querypersonarow{Medical Information}
1 & breast lump clinical evaluation
  & when to see doctor breast symptoms \\
2 & breast cancer symptoms clinical meaning
  & how doctors check breast lump \\
3 & mammogram vs ultrasound breast lump
  & breast MRI procedure explained \\
4 & breast biopsy what to expect
  & signs of breast cancer mammogram \\
5 & breast calcifications meaning
  & what doctors look for breast exam \\
6 & diagnosing breast cancer tests
  & breast imaging after symptoms \\
7 & core needle biopsy breast
  & breast surgeon consultation questions \\

\querypersonarow{Alternative Medicine}
1 & breast lump natural remedy
  & breast pain natural causes \\
2 & breast cyst natural treatment
  & holistic breast health tips \\
3 & breast swelling herbal remedies
  & natural breast health supplements \\
4 & breast tenderness without doctor
  & avoid biopsy breast lump \\
5 & breast symptoms home remedies
  & doctors overreact to breast lumps \\
6 & breast lump massage natural
  & breast health herbs benefits \\
7 & breast pain holistic approach
  & natural ways to reduce breast lump \\

\querypersonarow{Peer Narrative}
1 & breast lump stories from others
  & noticed breast change not sure \\
2 & breast symptom confusion stories
  & has anyone felt this breast \\
3 & breast change and denial stories
  & others with breast lump experience \\
4 & breast feeling weird same here
  & noticing breast changes and waiting \\
5 & breast symptom uncertainty stories
  & people who ignored breast changes \\
6 & breast lump maybe nothing stories
  & feeling unsure about breast change \\
7 & breast symptom shared experiences
  & what others noticed in breast \\

\end{tabular}
\end{ruledtabular}
\end{table*}

\begin{table*}[t]
\caption{TikTok search queries used in the Active Treatment context.}
\label{tab:active_treatment_queries}
\footnotesize
\renewcommand{\arraystretch}{1.12}
\setlength{\tabcolsep}{6pt}
\begin{ruledtabular}
\begin{tabular}{
    c
    p{0.39\textwidth}
    p{0.39\textwidth}
}
\textbf{Day}
& \multicolumn{1}{c}{\textbf{Query 1}}
& \multicolumn{1}{c}{\textbf{Query 2}} \\

\querypersonarow{Medical Information}
1 & breast cancer chemo side effects
  & taxol neuropathy management \\
2 & tamoxifen side effects management
  & ovarian suppression side effects \\
3 & radiation skin care breast
  & breast radiation fatigue relief \\
4 & port care during chemotherapy
  & chemo infusion what to expect \\
5 & sentinel lymph node biopsy recovery
  & post mastectomy drain care \\
6 & hormone therapy breast cancer guidelines
  & bone pain from breast cancer treatment \\
7 & breast cancer treatment timeline
  & evidence based breast cancer recovery \\

\querypersonarow{Alternative Medicine}
1 & natural breast cancer treatment
  & breast cancer alternative healing \\
2 & refused chemo breast cancer
  & breast cancer remission without chemo \\
3 & chemo alternatives breast cancer
  & holistic breast cancer recovery \\
4 & natural ways shrink breast tumor
  & integrative breast cancer treatment \\
5 & breast cancer without radiation
  & tamoxifen alternatives breast cancer \\
6 & breast cancer herbal remedies
  & alternative healing for breast cancer \\
7 & breast cancer recovery without chemo
  & natural breast tumor shrinkage \\

\querypersonarow{Peer Narrative}
1 & breast cancer chemo journey
  & breast cancer treatment stories \\
2 & how chemo felt breast cancer
  & radiation day in my life \\
3 & breast cancer support community
  & women sharing chemo stories \\
4 & breast cancer diagnosis feelings
  & coping emotionally during chemo \\
5 & breast cancer not alone
  & what treatment felt like \\
6 & breast cancer survivor story time
  & comfort after breast cancer diagnosis \\
7 & going through breast cancer together
  & emotional support during radiation \\

\end{tabular}
\end{ruledtabular}
\end{table*}

\section{Crossed-Random-Intercept Robustness Model}
\label{app:glmm}

As a robustness check, we fit a Bayesian variational logistic mixed model to
the cancer-relevant results. The model used the same framing-by-context fixed
effects as the primary regression and included crossed random intercepts for
account and query.

\begin{table*}[t]
\centering
\footnotesize
\caption{Fixed-effect estimates from the variational-Bayes logistic mixed
model predicting possible misinformation among cancer-relevant results. The
model includes crossed random intercepts for account and query. Medical
Information in the Symptom Noticing context is the reference condition.
Intervals are approximate 95\% posterior intervals for the odds ratios.}
\label{tab:glmm}

\begin{ruledtabular}
\begin{tabular}{@{}lrrr@{}}
\textbf{Predictor} & \textbf{Posterior mean $\hat\beta$} &
\textbf{OR} & \textbf{Approx.\ 95\% OR interval} \\
\hline
Intercept (Medical Information, Symptom Noticing)
  & $-$3.08 & 0.05 & [0.04, 0.05] \\
Alternative Medicine
  & 3.07 & 21.49 & [19.66, 23.49] \\
Peer Narrative
  & $-$0.51 & 0.60 & [0.47, 0.76] \\
Active Treatment
  & 0.00 & 1.00 & [0.91, 1.10] \\
Alternative Medicine $\times$ Active Treatment
  & 0.28 & 1.32 & [1.18, 1.47] \\
Peer Narrative $\times$ Active Treatment
  & $-$0.47 & 0.62 & [0.44, 0.89] \\
\end{tabular}
\end{ruledtabular}
\end{table*}

The estimated random-intercept standard deviations were 0.09 for account and
1.14 for query. The model reproduced the large positive Alternative Medicine contrast
and negative Peer Narrative contrast. Because variational-Bayes uncertainty
can be too narrow, these estimates are used only as a directional robustness
check. Statistical inference is based on the primary two-way cluster-robust
logistic regression.

\newpage
\section{Supplementary Results}
\label{app:supp-results}

\subsection{Misinformation Rates by Condition}

\begin{figure}[t]
    \centering
    \includegraphics[width=\linewidth]{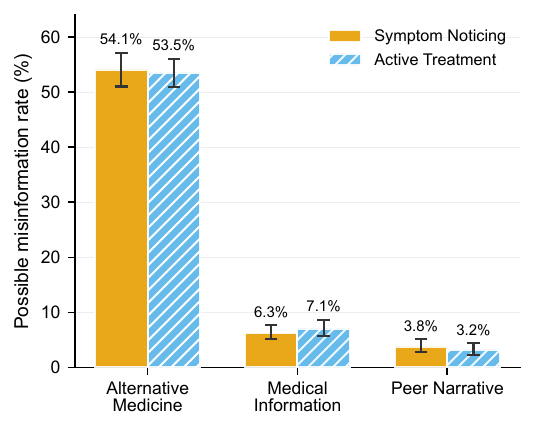}
    \caption{Possible misinformation rates by query framing and search context
    among cancer-relevant search results. Error bars show 95\% Wilson confidence
    intervals.}
    \label{fig:supp-rates}
\end{figure}

Figure~\ref{fig:supp-rates} presents the misinformation rates for all six
combinations of query framing and search context. Alternative Medicine queries
returned possible misinformation in 54.1\% of cancer-relevant results in the
Symptom Noticing context and 53.5\% in the Active Treatment context. The
corresponding rates were 6.3\% and 7.1\% for Medical Information and 3.8\% and
3.2\% for Peer Narrative. The large difference associated with query framing
was therefore observed in both search contexts.

\subsection{Variation Across Queries and Accounts}

\begin{figure*}[p]
    \centering
    \includegraphics[width=\linewidth]{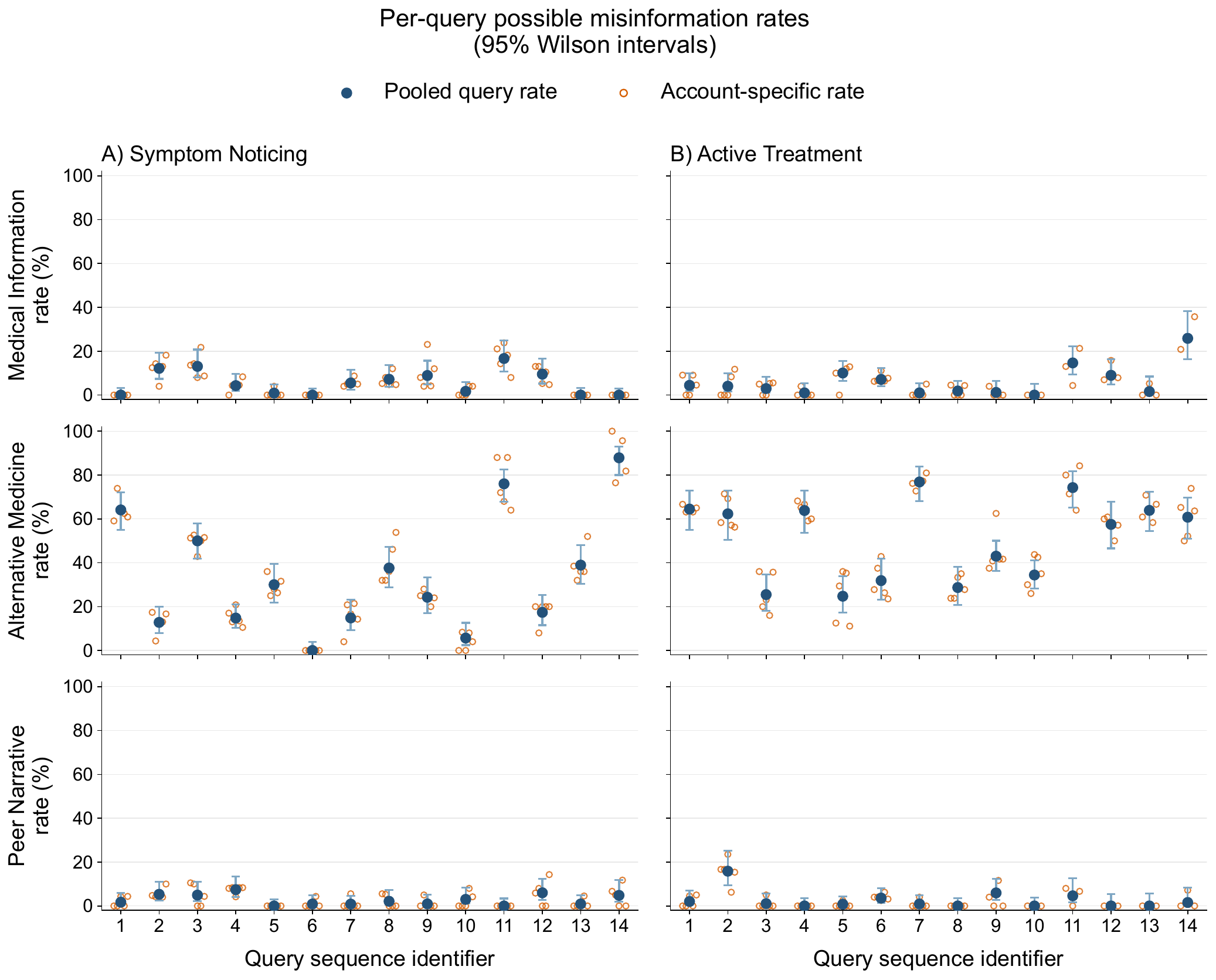}
    \caption{Per-query possible misinformation rates among all search
    results. Rows represent query framings and columns represent search
    contexts. Filled circles show the rate pooled across accounts for each
    query, with 95\% Wilson confidence intervals. Hollow circles show
    account-specific rates. Query sequence identifiers distinguish the 14
    queries within each condition; they do not represent result rank, and the
    same identifier across panels does not indicate the same query.}
    \label{fig:supp-per-query}
\end{figure*}

Condition-level averages conceal substantial variation among individual queries
(Figure~\ref{fig:supp-per-query}). Among all search results, pooled
per-query misinformation rates for Alternative Medicine ranged from 0\% to
87.9\% in the Symptom Noticing context and from 24.8\% to 76.9\% in the Active
Treatment context. The corresponding ranges were 0\%--16.7\% and 0\%--25.9\%
for Medical Information, and 0\%--7.4\% and 0\%--15.9\% for Peer Narrative.
The account-specific estimates also show variation across accounts receiving
the same query. Despite this variation, most Alternative Medicine queries
returned substantially higher misinformation rates than Medical Information
and Peer Narrative queries.

\subsection{Distribution of Misinformation Types}

\begin{figure*}[t]
    \centering
    \includegraphics[width=\linewidth]{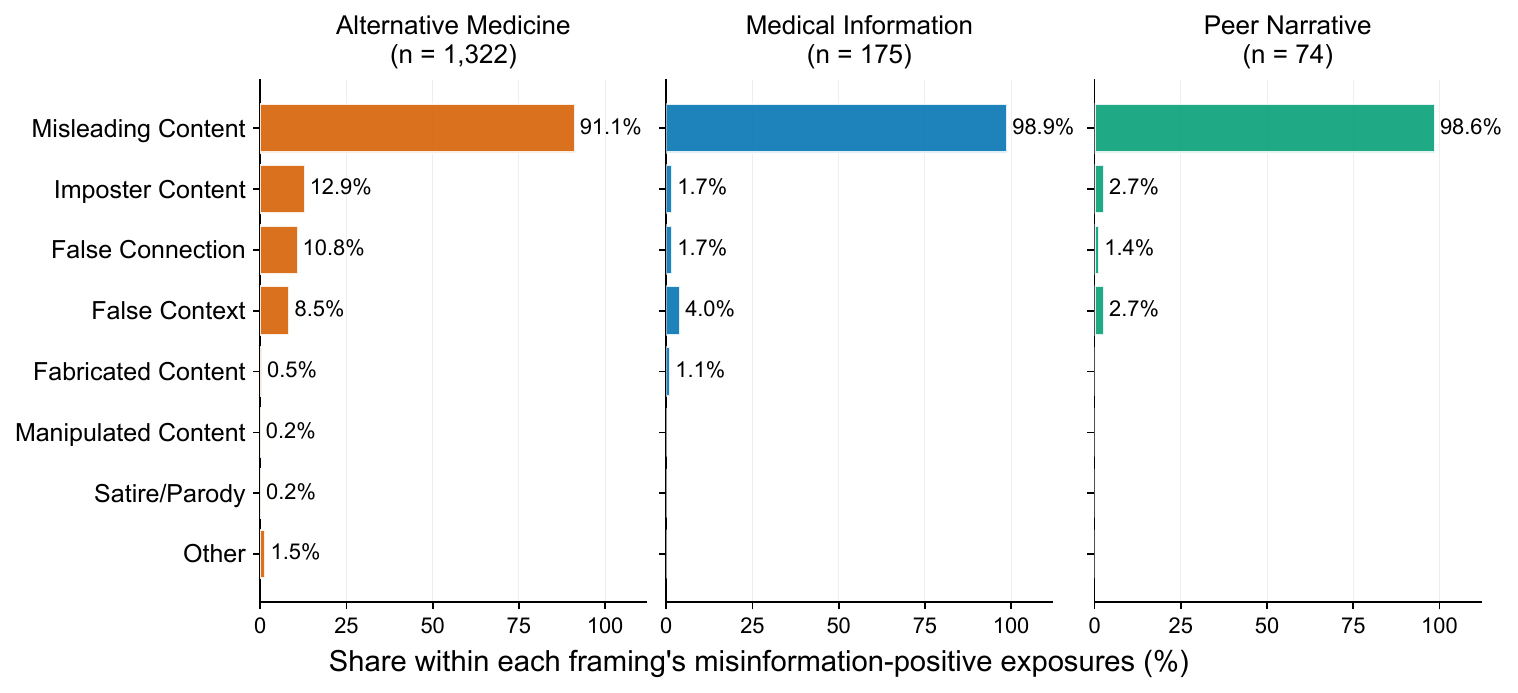}
    \caption{Distribution of misinformation types by query framing among
    cancer-relevant results classified as containing possible misinformation,
    pooling the Symptom Noticing and Active Treatment contexts. Each panel uses
    the misinformation-positive results within that framing as its denominator
    (Alternative Medicine, $n=1{,}322$; Medical Information, $n=175$; Peer
    Narrative, $n=74$). A result could receive more than one type; consequently,
    percentages within a panel may sum to more than 100\%.}
    \label{fig:supp-types}
\end{figure*}

Figure~\ref{fig:supp-types} shows the misinformation types assigned to the
1,571 cancer-relevant results classified as containing possible misinformation.
Misleading Content was the dominant type under every query framing: 91.1\% for
Alternative Medicine, 98.9\% for Medical Information, and 98.6\% for Peer
Narrative. Alternative Medicine results also received Imposter Content
(12.9\%), False Connection (10.8\%), and False Context (8.5\%) labels more
frequently than results under the other two framings. These comparisons are
descriptive because the number of misinformation-positive results differs
substantially across the three query framings.

\end{document}